\documentclass[fleqn,usenatbib]{mnras}

\usepackage{newtxtext,newtxmath}

\usepackage[T1]{fontenc}
\usepackage{ae,aecompl}

\usepackage{graphicx}	
\usepackage{amsmath}	
\usepackage{xcolor} 
\usepackage[normalem]{ulem} 

\title[The orbit of WR\,140]{The orbit and stellar masses of the archetype colliding-wind binary WR\,140}
\author[J. D. Thomas et al.]{Joshua D. Thomas$^{1}$\thanks{E-mail: jthomas@clarkson.edu},
Noel D. Richardson$^{2}$\thanks{E-mail: noel.richardson@erau.edu},
J. J. Eldridge$^{3}$,
Gail H. Schaefer$^{4}$,
\newauthor
John D. Monnier$^{5}$,
Hugues Sana$^{6}$,
Anthony F. J. Moffat$^{7}$,
Peredur Williams$^{8}$,
\newauthor
Michael F. Corcoran$^{9,10}$,
Ian R. Stevens$^{11}$,
Gerd Weigelt$^{12}$,
Farrah D. Zainol$^{11}$,
\newauthor
Narsireddy Anugu$^{13,14}$,
Jean-Baptiste Le Bouquin$^{15}$,
Theo ten Brummelaar$^{4}$,
\newauthor
Fran Campos$^{16}$,
Andrew Couperus$^{1,17}$,
Claire L. Davies$^{13}$,
Jacob Ennis$^{5}$,
\newauthor
Thomas Eversberg$^{18}$,
Oliver Garde$^{19}$,
Tyler Gardner$^{5}$,
Joan Guarro Fl\'{o}$^{20}$,
\newauthor
Stefan Kraus$^{13}$,
Aaron Labdon$^{13}$,
Cyprien Lanthermann$^{6,15}$,
Robin Leadbeater$^{21}$,
\newauthor
T. Lester$^{22}$,
Courtney Maki$^{1}$,
Brendan McBride$^{1}$,
Dogus Ozuyar$^{23}$,
J. Ribeiro$^{24}$,
\newauthor
Benjamin Setterholm$^{5}$,
Berthold Stober$^{25}$,
Mackenna Wood$^{1}$,
Uwe Zurm\"uhl$^{26}$
\\
$^{1}$Department of Physics, Clarkson University, 8 Clarkson Ave, Potsdam, NY 13699, USA \\
$^{2}$Department of Physics and Astronomy, Embry-Riddle Aeronautical University, 3700 Willow Creek Road, Prescott, AZ 86301, USA\\
$^{3}$Department of Physics, University of Auckland, Private Bag 92019, Auckland 1010, New Zealand\\
$^{4}$The CHARA Array of Georgia State University, Mount Wilson Observatory, Mount Wilson, CA 91023, USA\\
$^{5}$Department of Astronomy, University of Michigan, 1085 S. University, Ann Arbor, Michigan 48109, USA\\
$^{6}$Institute of Astrophysics, KU Leuven, Celestijnlaan 200D, 3001, Leuven, Belgium\\
$^{7}$Centre de Recherche en Astrophysique du Qu\'ebec, D\'epartement de physique, Universit\'e de Montr\'eal,\\ CP 6128, Succ. C.-V., Montr\'eal, QC H3C 3J7\\
$^{8}$Institute for Astronomy, University of Edinburgh, Royal Observatory, Blackford Hill,Edinburgh EH9 3HJ\\
$^{9}$CRESST II \& X-ray Astrophysics Laboratory, Code 662, NASA Goddard Space Flight Center, Greenbelt, MD 20771, USA\\
$^{10}$Institute for Astrophysics and Computational Sciences, Department of Physics, The Catholic University of America, \\Washington, DC 20064, USA\\
$^{11}$School of Physics and Astronomy, University of Birmingham, Edgbaston, Birmingham B15 2TT, UK \\
$^{12}$Max Planck Institute for Radio Astronomy, Auf dem H\"{u}gel 69, D-53121 Bonn, Germany\\
$^{13}$University of Exeter, Department of Physics and Astronomy, Exeter, Devon EX4 4QL, UK\\
$^{14}$Steward Observatory, 933 N. Cherry Avenue, University of Arizona, Tucson, AZ 85721, USA\\
$^{15}$Institut de Plan\'etologie et d'Astrophysique de Grenoble, France\\
$^{16}$Observatori Puig d'Agulles, Passatge Bosc 1, 08759, Vallirana, Barcelona, Spain\\
$^{17}$Department of Physics and Astronomy, Georgia State University, 33 Gilmer Street SE Atlanta, GA 30303\\
$^{18}$Schn\"{o}rringen Telescope Science Institute, Waldbr\"{o}l, Germany\\
$^{19}$Observatoire de la Tourbi\'ere, 38690 CHABONS, France\\
$^{20}$Balmes 2, 08784 Piera, Barcelona, Spain\\
$^{21}$The Birches Torpenhow, Wigton, Cumbria CA7 1JF, UK\\
$^{22}$1178 Mill Ridge Road, Arnprior, ON, K7S3G8, Canada\\
$^{23}$Ankara University, Faculty of Science, Dept. of Astronomy and Space Sciences, 06100, Tandogan, Ankara, Turkey\\
$^{24}$Observat\'orio do Centro de Informa\c{c}\~{a}o Geoespacial do Ex\'ercito - Lisboa, Portugal\\
$^{25}$VdS Section Spectroscopy, Germany; Teide Pro-Am Collaboration\\
$^{26}$D-31180 Giesen, Lower Saxony, Germany
}

\date{Accepted XXX. Received YYY; in original form ZZZ}

\pubyear{2021}

\begin{document}
\label{firstpage}
\pagerange{\pageref{firstpage}--\pageref{lastpage}}
\maketitle
\clearpage
\begin{abstract}
We present updated orbital elements for the Wolf-Rayet (WR) binary WR\,140 (HD\,193793; WC7pd + O5.5fc). The new orbital elements were derived using previously published measurements along with {\color{black}160} new radial velocity measurements across the 2016 periastron passage of WR\,140. Additionally, four new measurements of the orbital astrometry were collected with the CHARA Array. With these measurements, we derive stellar masses of
$M_{\rm WR} = 10.31\pm0.45 M_\odot$ and $M_{\rm O} = 29.27\pm1.14 M_{\odot}$.
We also include a discussion of the evolutionary history of this system from the Binary Population and Spectral Synthesis (BPASS) model grid to show that this WR star likely formed primarily through mass loss in the stellar winds, with only a moderate amount of mass lost or transferred through binary interactions.
\end{abstract}




\section{Introduction}

Mass is the most fundamental property of a star, as it constrains most properties of its evolution. Accurate stellar mass determinations are therefore critical to test stellar evolutionary models and to measure the effects of binary interactions. So far, only two carbon-rich Wolf-Rayet (WR) stars have established visual and double-lined spectroscopic orbits, the hallmark of mass measurements.  They are $\gamma^2$\,Velorum (WC8+O7.5III-V) \citep{2007MNRAS.377..415N,2017MNRAS.468.2655L, 2017MNRAS.471.2715R}  and WR\,140 \citep{2011MNRAS.418....2F, 2011ApJ...742L...1M}.

$\gamma^2$\,Vel contains the closest WR star to us at 336 pc \citep{2017MNRAS.468.2655L}, allowing interferometry to resolve the close 78-d orbit. The only other WR system with a reported visual orbit is WR\,140 \citep{2011ApJ...742L...1M}, a long-period highly eccentric system and a benchmark for massive colliding-wind systems, and the subject of this paper. {\color{black} The only nitrogen-rich WR binary with a resolved orbit is WR\,133, which was recently reported by Richardson et al.~(2021).} Some progress has also been made in increasing this sample by \citet{2016MNRAS.461.4115R}, who resolved the long-period systems WR\,137 and WR\,138 with the CHARA Array.

WR\,140 is a very intriguing object; with a long period (P=7.992 years) and a high eccentricity ($e = 0.8993$), the system has some resemblance to the enigmatic massive binary $\eta$ Carinae. It has a double-lined spectroscopic and visual orbit, meaning that we possess exceptional constraints on the system's geometry at any epoch.

WR\,140 was one of the first WC stars found to exhibit infrared variability attributed to dust formation \citep{1978MNRAS.185..467W}. Its radio, and X-ray emissions, along with the dusty outbursts in the infrared, were originally proposed to be modulated by its binary orbit by \citet{1990MNRAS.243..662W}. \citet{2009MNRAS.395.1749W} showed that dust production was indeed modulated by the elliptical orbit. Recently, \citet{2020ApJ...898...74L} showed that WC binaries with longer orbital periods produced larger dust grains than shorter period systems. Therefore, the accurate determination of all related properties of these binaries can help test this trend, and provide critical constraints on mechanisms that produce dust in these systems.

The orbital properties and apparent brightness of WR\,140 make it an important system for the study of binary evolution.  As one of the few Wolf-Rayet stars with an exceptionally well-determined orbit, it serves as an important astrophysical laboratory for dust production \citep[e.g.,][]{2009MNRAS.395.1749W} and colliding-wind shock physics \citep[e.g.,][]{2015PASJ...67..121S}. In this paper we present refined orbital parameters based on new interferometric and spectroscopic measurements focused on the December 2016 periastron passage. Section 2 presents the observations.  We present our new orbital elements and masses in Section 3, and then discuss the evolutionary history of WR\,140 in Section 4.  We summarize our findings in Section 5.

\section{Observations}
\subsection{Spectroscopic Observations}

During the 2016 periastron passage of WR\,140, we initiated a global spectroscopic campaign on the system similar to that described by \citet{2011MNRAS.418....2F}. In total, our {\color{black}Pro-Am campaign} collected {\color{black}160} spectra over 323 days when the velocities were expected to be varying most rapidly. Our measurements are provided in the appendix of this paper in Table~\ref{RadialVelocities}.
The spectra all covered the C\,{\sc iii} $\lambda 5696\text{\normalfont\AA}$ emission-line (broad and narrow components emitted in the WR- and O star winds, respectively, and from the variable CW region) and the He\,{\sc i} $\lambda 5876\text{\normalfont\AA}$ line (with emission and P Cygni absorption components from the WR wind, a variable excess emission from the colliding-wind shock-cone, and an absorption component from the O star's photosphere).  {\color{black}In addition we downloaded archival ESPaDOnS spectra\footnote{http://polarbase.irap.omp.eu/} \citep{1997MNRAS.291..658D,2014PASP..126..469P}, and previously analyzed by \citet{2014ApJ...781...73D}.  There were a total of 6 nights of data that were co-added to make a single spectrum for each night. }

\subsubsection{Radial Velocity Measurements}

\begin{figure}
\begin{center}
\includegraphics[angle=0, width=8cm]{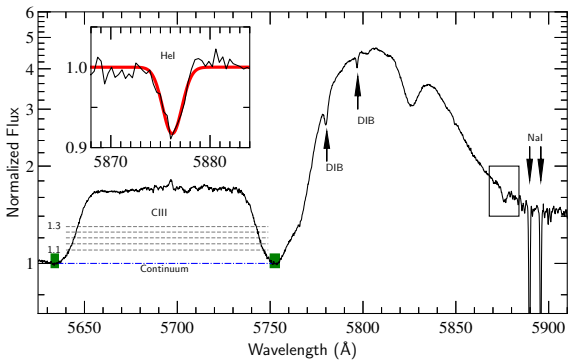}
\end{center}
\caption{{\color{black} An example spectrum, collected on HJD 2457703.3, with annotations to illustrate the measurement process. On the left of is the flat-topped C\,{\sc iii} line profile used for determining the radial velocity of the WR component of the system.   The grey dashed lines correspond to the five normalized flux values use for bisection, with the extreme values marked to the left of the grey dashed lines. The small central peak near 5700\,$\text{\normalfont\AA}$ is the C\,{\sc iii} component from the O star.  The blue dot-dash line is the continuum, and the green regions on either side of the C\,{\sc iii} profile contain the regions used for normalizing the feature.  Two upward arrows indicate the locations of the DIBs used to check the wavelength calibration.  The two downward arrows mark the interstellar Na\,{\sc i} D-lines, also used to check the wavelength solution.  The inset illustrates the normalized He\,{\sc i} line from the O star, denoted by the box next to the Na\,{\sc i} lines.  The red curve is a Gaussian fit to the O star.
}}
\label{examplespec}
\end{figure}

The properties of the spectra, and a journal of the observations, are shown in Table~\ref{specobs-table}. With spectra from so many different sources, we had to ensure that the wavelength calibration was reliable among the various observatories.  We therefore checked the alignment of the interstellar Na D absorption lines {\color{black} and Diffuse Interstellar Bands (DIBs) with locations indicated in Fig.\ \ref{examplespec} and wavelengths measured in the ESPaDOnS data.   We then linearly shifted the data by no more than $1.3 \text{\normalfont\AA}$ to obtain a better wavelength solution. With {\color{black}four interstellar} absorption lines, we were also able to ensure that the spectral dispersion was reliable for the data during this process. An example spectrum of the C\,{\sc iii} $\lambda 5696\text{\normalfont\AA}$ line is shown in Fig.~\ref{examplespec}. }

\begin{table*}
\begin{minipage}{160mm}
\centering
\caption{List of contributed spectra, in order of number of spectra.  The wavelength coverage and range of observation data for each primary observer are noted, as well as the approximate resolving power of their spectra. {\color{black} The average signal to noise ratio for each observer is also noted.}
\label{specobs-table}}
\begin{tabular}{l c c c c c c c c c}
\hline \hline
Observer	&	 $N_{\rm spectra}$ 	&	 $\lambda_{\rm start}$	&	$\lambda_{\rm end}$ 	&	HJD$_{\rm first}$	&	 HJD$_{\rm end}$	&	 Resolving	&	Average	&	Spectrograph	&	Aperature	\\
 	&	 	&	(\AA)	&	(\AA)	&	$-2450000.5$	&	$-2450000.5$	&	Power	&	S/N	&		&	(m)	\\ \hline
Guarro	&	48	&	3979	&	7497	&	7666.89	&	7944.85	&	\phantom{1}9,000	&	100	&	eSHEL	&	0.4	\\
Thomas	&	26	&	5567	&	6048	&	7644.12	&	7918.07	&	\phantom{1}5,000	&	50	&	LHIRES III	&	0.3	\\
Leadbeater	&	17	&	5623	&	5968	&	7615.9	&	7788.73	&	\phantom{1}5,000	&	173	&	LHIRES III	&	0.28	\\
Ribeiro	&	16	&	5528	&	6099	&	7709.81	&	7762.76	&	\phantom{1}6,000	&	70	&	LHIRES III	&	0.36	\\
Garde	&	10	&	4185	&	7314	&	7624.91	&	7759.69	&	11,000	&	83	&	eSHEL	&	0.3	\\
Berardi	&	12	&	5522	&	6002	&	7715.73	&	7778.71	&	\phantom{1}5,000	&	180	&	LHIRES III	&	0.23	\\
Campos	&	12	&	5463	&	6212	&	7675.86	&	7764.73	&	\phantom{1}5,000	&	65	&	DADOS	&	0.36	\\
Lester	&	9	&	5143	&	6276	&	7697.01	&	7769.94	&	\phantom{1}7,000	&	118	&	LHIRES III	&	0.3	\\
Ozuyar	&	6	&	4400	&	7397	&	7624.77	&	7730.68	&	\phantom{1}2,000	&	85	&	eSHEL	&	0.4	\\
{\color{black}ESPaDOnS }	&	6	&	3691	&	10481	&	4645.59	&	8293.62	&	\phantom{8}1,000	&	191	&	ESPaDOnS	&	3.58	\\
Stober	&	1	&	4276	&	7111	&	7616.82	&	--	&	\phantom{1}8,000	&	36	&	eSHEL	& 0.3		\\

\hline \hline
\end{tabular}
\end{minipage}
\end{table*}

\begin{figure*}
\begin{center}
\includegraphics[angle=0, width=8cm]{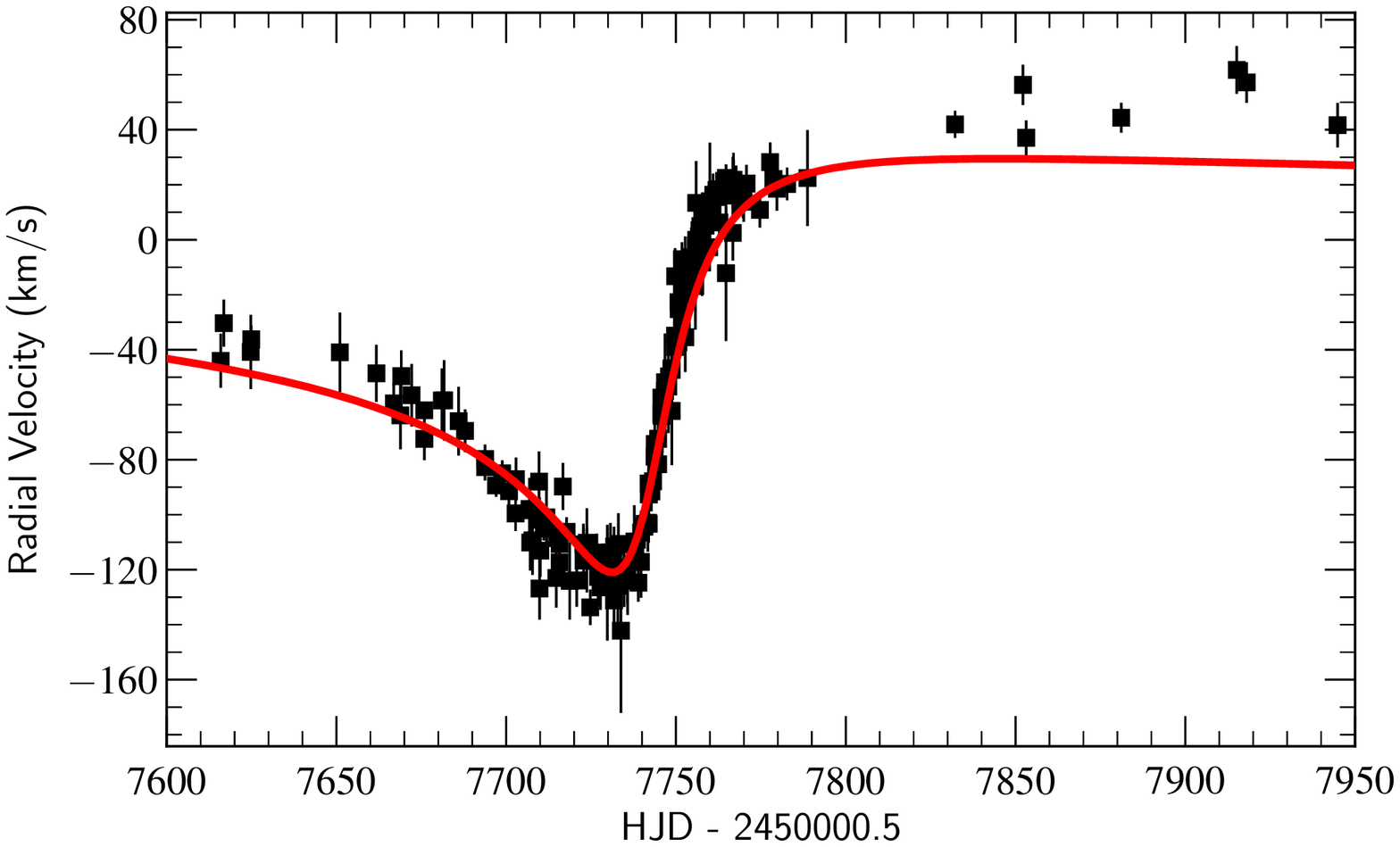}
\includegraphics[angle=0, width=8cm]{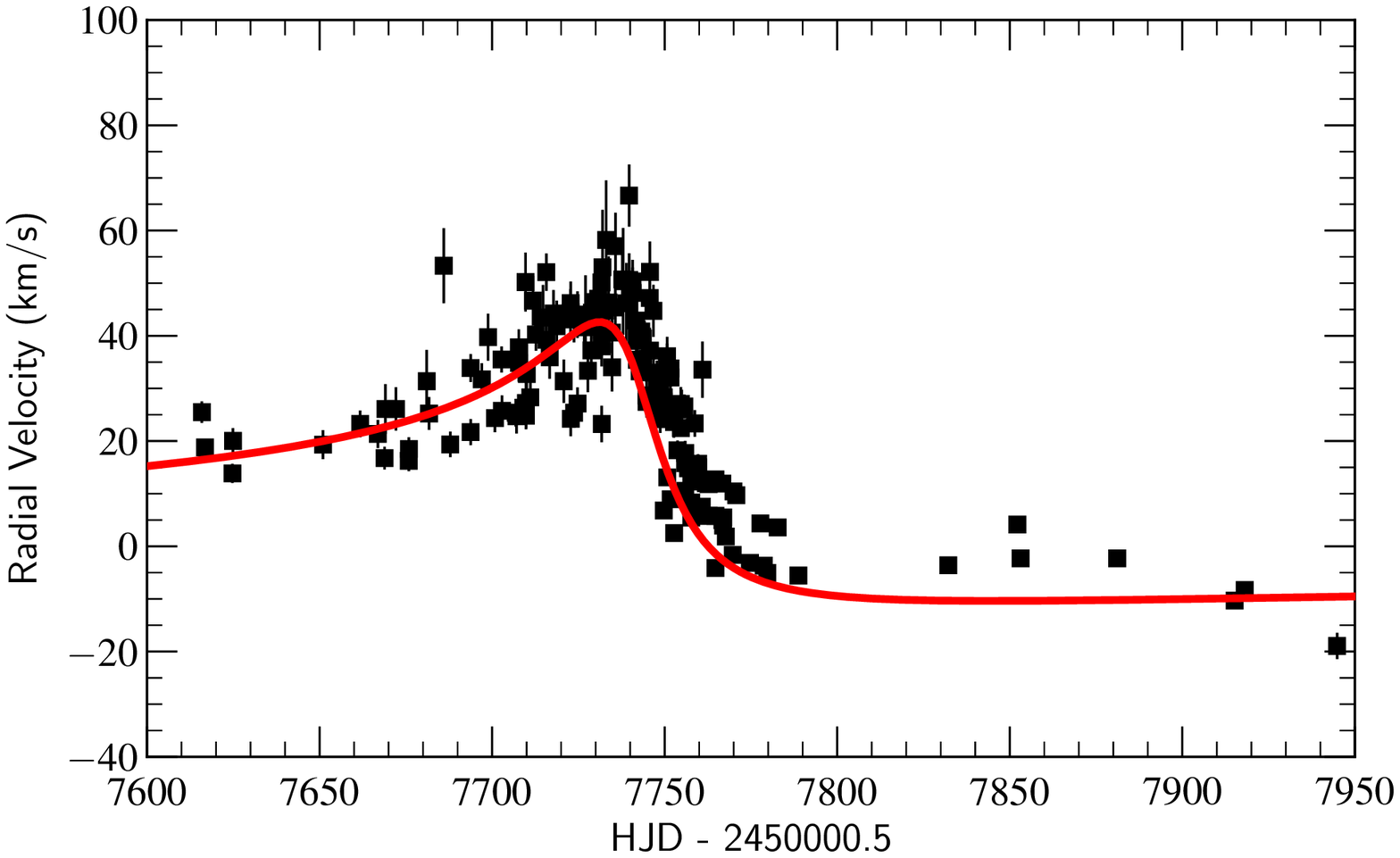}
\end{center}
\caption{The left panel contains the measured radial velocities from the 2016 periastron passage for the WR star.  The right panel shows the measured radial velocities for the O star companion. {\color{black}The error bars in both panels are discussed in the text.}  The red curves plotted here corresponds to the orbital elements reported in this paper. }
\label{measuredRV}
\end{figure*}

The velocities of the WR~star, shown in the left panel of Fig.~\ref{measuredRV}, were measured by bisecting the C\,{\sc iii} $5696\text{\normalfont\AA}$ emission plateau to find the centroid of the feature. We chose this line due to its relative isolation from other emission features. For example, the C\,{\sc iv} $\lambda\lambda 5802,5812\text{\normalfont\AA}$ doublet may have been a better choice, but is heavily blended with the He\,{\sc i} $\lambda 5876\text{\normalfont\AA}$ emission from the WR wind. The spectra were normalized with a linear function so that the low points on either side of the C\,{\sc iii} feature had a flux of unity. The emission profile was bisected at normalized flux values, illustrated in Fig.~\ref{examplespec}, of: 1.1, 1.15, 1.2, 1.25, and 1.3. The velocity was then calculated for the average bisector. The displayed error bars take into account the standard deviation in the bisection velocity, the signal-to-noise in the continuum regions selected for the normalization, and the wavelength calibration.  The errors were added in quadrature.  It was found that the error is dominated by the standard deviation in the bisectors.

A few velocity measurements made just post HJD~2457800 do seem higher than expected for a Keplerian orbit, close examination of these spectra reveal that the colliding-wind excess is likely affecting the red shoulder of the C\,{\sc iii} emission profile and skews the bisector toward higher redshift in our measurements.  The variation in the location of the red shoulder corresponds to skew in the bisector of approximately 30 km s$^{-1}$, which is roughly the difference between the outliers and the model fit. We did not attempt to correct this, as the number of points affected was small, and this phase of the binary orbit has minimal changes in the radial velocity.

The O star velocities in the right panel of Fig.~\ref{measuredRV} were measured by fitting a {\color{black}Gaussian} profile to the He\,{\sc i} $\lambda5875.621 \text{\normalfont\AA}$ helium absorption line, which never interferes with any P Cygni absorption from the WR star due to the high WR wind speed. When phase-folded, our O star velocities are consistent with velocities from a large range of absorption lines measured by \citet{2011MNRAS.418....2F}. {\color{black}The displayed error bars for the O star velocities account for the uncertainty in the wavelength calibration, the standard deviation, and the uncertainty in the centroid of a Gaussian.  We used the FWHM from our Gaussian profile in equation 15 of \citet{1987NIMPB..28..146G} to find the uncertainty in the centroid.  The reported error is the quadrature sum of the errors.}  We found that the largest source of uncertainty in the centroid of the Gaussian fit was caused by the signal-to-noise in the continuum.

\subsection{Interferometry with the CHARA Array}

We have obtained four new epochs of CHARA Array interferometry to measure the precise astrometry of the component stars, following the work of \citet{2011ApJ...742L...1M}. The first observation was obtained on 2011 June 17 with the CLIMB beam combiner \citep{2013JAI.....240004T}. This observation consisted of five observations with the E1, W1, and W2 telescopes. Observations were calibrated with the same calibration stars as \citet{2011ApJ...742L...1M}, with the observations of the calibration stars happening before and after each individual scan. These bracketed observations were made in the $K'-$band and reduced with a pipeline written by John D. Monnier, and were then combined into one measurement to improve the astrometric accuracy.

Another observation was obtained with the MIRC combiner \citep{2012SPIE.8445E..0YM} on UT 2011 September 16.  The MIRC combiner uses all six telescopes of the CHARA Array, with eight spectral channels across the $H-$band.  The data were reduced using the MIRC data reduction pipeline \citep{2007Sci...317..342M} using a coherent integration time of 17~ms. {\color{black} \citet{2012ApJ...761L...3M} determined a correction factor for the absolute wavelength scale of MIRC data by analyzing the orbit of $\iota$ Peg. Based on that analysis, we multiplied the wavelengths in the calibrated data file by a factor of 1.004, appropriate for 6-telescope MIRC data collected between 2011-2017. Therefore, }
 we applied this wavelength correction factor of 1.004 to the data based on the analysis by \citet{2012ApJ...761L...3M}. Two additional observations were obtained with the upgraded MIRC-X combiner \citep{2018SPIE10701E..23K, 2018SPIE10701E..24A, 2020arXiv200712320A} on UT~2018~October~26 and 2019~July~1.  The observations were recorded in the PRISM50 mode which provides a spectral resolution of $R = 50$.  The data were reduced using the MIRC-X data reduction pipeline, version 1.2.0\footnote{https://gitlab.chara.gsu.edu/lebouquj/mircx\_pipeline.git.} to produce calibrated visibilities and closure phases.  During the reduction, we applied the bias correction included in the pipeline and set the number of coherent coadds to 5.  A list of the calibrators and their angular diameters ($\theta_{\rm UD}$) adopted from the JMMC catalog \citep{2017yCat.2346....0B} are listed in Table~\ref{calibrators}.

We analyzed the calibrated interferometric data using the same approach as \citet{2016MNRAS.461.4115R}.  More specifically, we performed an adaptive grid search to find the best fit binary position and flux ratio ($f_{\rm WR}/f_{\rm O}$) using software\footnote{This software is available at http://www.chara.gsu.edu/analysis-software/binary-grid-search.} developed by \citet{2016AJ....152..213S}.  During the binary fit, we fixed the uniform disk diameters of the components to sizes of 0.05~mas for the WR star and 0.07 mas for the O star as determined by \citet{2011ApJ...742L...1M}.  We added a contribution from excess, over-resolved flux to the binary fit that varied during each epoch.  The uncertainties in the binary fit were derived by adding in quadrature errors computed from three sources: the formal covariance matrix from the binary fit, the variation in parameters when changing the coherent integration time used to reduce the data (17~ms and 75~ms for MIRC; 5 and 10 coherent coadds for MIRC-X), and the variation in parameters when changing the wavelength scale by the internal precision (0.25\% for MIRC determined by \citet{2011ApJ...742L...1M}; 0.5\% for MIRC-X determined by \citet{2020arXiv200712320A}). In scaling the uncertainties in the position, we added the three values in quadrature for the major axis of the error ellipse ($\sigma_{\rm major}$) and scaled the minor axis ($\sigma_{\rm minor}$) to keep the axis ratio and position angle fixed according to the values derived from the covariance matrix. The results of the astrometric measurements are given in Table~\ref{interferometry}, with significant figures dependent on the individual measurements. In addition to the previously discussed parameters, we include the position angle of the error ellipse ($\sigma_{\rm PA}$) in Table~\ref{interferometry}.

\begin{table*}
\begin{minipage}{170mm}
\centering
\caption{Calibrator stars observed during the MIRC and MIRC-X observations at the CHARA Array.
\label{calibrators}}
\begin{tabular}{l c c}
\hline \hline
Star       & $\theta_{\rm UD}$ (mas) & Date Observed \\
\hline
HD 178538  & 0.2487 $\pm$ 0.0062 & 2019Jul01 \\
HD 191703  & 0.2185 $\pm$ 0.0055 & 2019Jul01 \\
HD 197176  & 0.2415 $\pm$ 0.0058 & 2019Jul01 \\
HD 201614  & 0.3174 $\pm$ 0.0074 & 2019Jul01 \\
HD 204050  & 0.4217 $\pm$ 0.0095 & 2018Oct26 \\
HD 228852  & 0.5441 $\pm$ 0.0127 & 2018Oct26 \\
HD 182564  & 0.3949 $\pm$ 0.0253 & 2011Sep16 \\
HD 195556  & 0.2118 $\pm$ 0.0080 & 2011Sep16 \\
HD 210839  & 0.4200 $\pm$ 0.0200 & 2011Sep16 \\
HD 214734  & 0.3149 $\pm$ 0.0286 & 2011Sep16 \\
\hline \hline
\end{tabular}
\end{minipage}
\end{table*}

\begin{table*}
\begin{minipage}{170mm}
\centering
\caption{Interferometric measurements with the CHARA Array.
\label{interferometry}}
\begin{tabular}{l l c c c c c c c c c}
\hline \hline
UT Date & HJD  & Instrument &   Bandpass    &   Separation  & Position  & $\sigma_{\rm major}$   & $\sigma_{\rm minor}$ & $\sigma_{\rm PA}$  & $f_{\rm WR}/f_{\rm O}$ & Excess Flux   \\
        &  $-2450000.5$     &            &               &   (mas) & Angle ($^\circ$)  &  (mas)               & (mas)                & ($^\circ$)         &                        & (\%)          \\ \hline
2011Jun17 &  5729.411 & CLIMB  & $K'$ &  13.02   &  153.00  &  0.22   &  0.06  & 162     &                   &                      \\

{\color{black}2011Sep16 }& 5820.270 & MIRC   & H & 12.969 & 151.749 & 0.065 & 0.049 & 111.65 & 1.5665 $\pm$ 0.2257 &  5.94 $\pm$ 0.81 \\
2018Oct26 & 8417.139 & MIRC-X & H & 11.932 & 155.969 & 0.060 & 0.043 & 141.12 & 1.1298 $\pm$ 0.0044 & 11.78 $\pm$ 0.12 \\
2019Jul01 & 8665.351 & MIRC-X & H & 13.017 & 152.458 & 0.065 & 0.029 & 173.43 & 1.1006 $\pm$ 0.0063 &  1.31 $\pm$ 0.77 \\

\hline  \hline
\end{tabular}
\end{minipage}
\end{table*}

\section{The Orbital Elements} \label{orbitalelements}
Orbital fits for massive stars with both high-quality spectroscopic and interferometric measurements have become more routine. For this work we simultaneously fit the spectroscopic and interferometric data using the method discussed in \citet{2016AJ....152..213S}, which was also used in Richardson et al.~(2021).  With the orbital solution from \citet{2011ApJ...742L...1M} as the starting point, the orbital models were simultaneously adjusted to fit radial velocities (from this work and \citet{2011MNRAS.418....2F}), and the interferometric measurements from this work, and from \citet{2011ApJ...742L...1M}. The models are adjusted to minimize the $\chi^2$ statistic.  We adopted a minimum 5~km~s$^{-1}$ error on the radial velocities so that the radial velocity and astrometric data have similar weight in the final $\chi^2$.

\begin{figure}
\begin{center}
\includegraphics[angle=0, width=8cm]{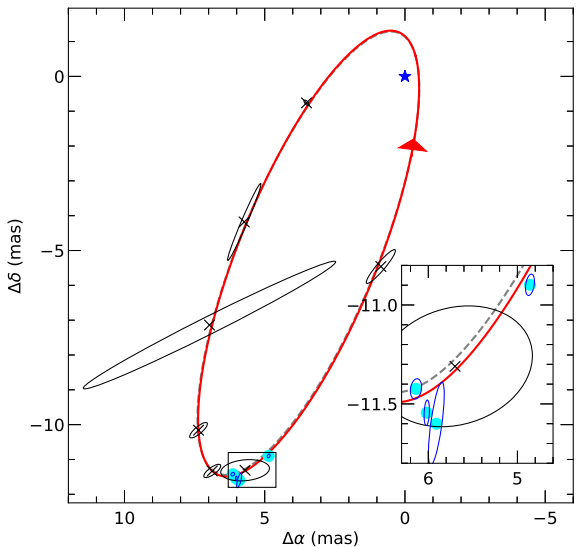}
\end{center}
\caption{The visual orbit with the O star positions relative to the WR star.  The WR star location is denoted by the blue star. The data from \citet{2011ApJ...742L...1M} are  shown with black $\times$ and their error ellipses.  The four new epochs of O star positions are shown as solid cyan circles.  The error ellipses on the new points are smaller than the symbol used.  The inset plot shows the error ellipses on the new CHARA data.  The solid red ellipse is the
{\color{black} fit} from this work.  The grey dashed ellipse is the best fit model from  \citet{2011ApJ...742L...1M} and the two solutions show remarkable agreement.}
\label{orbitfit}
\end{figure}

\begin{figure*}
\begin{center}
\includegraphics[angle=0, width=8.5cm]{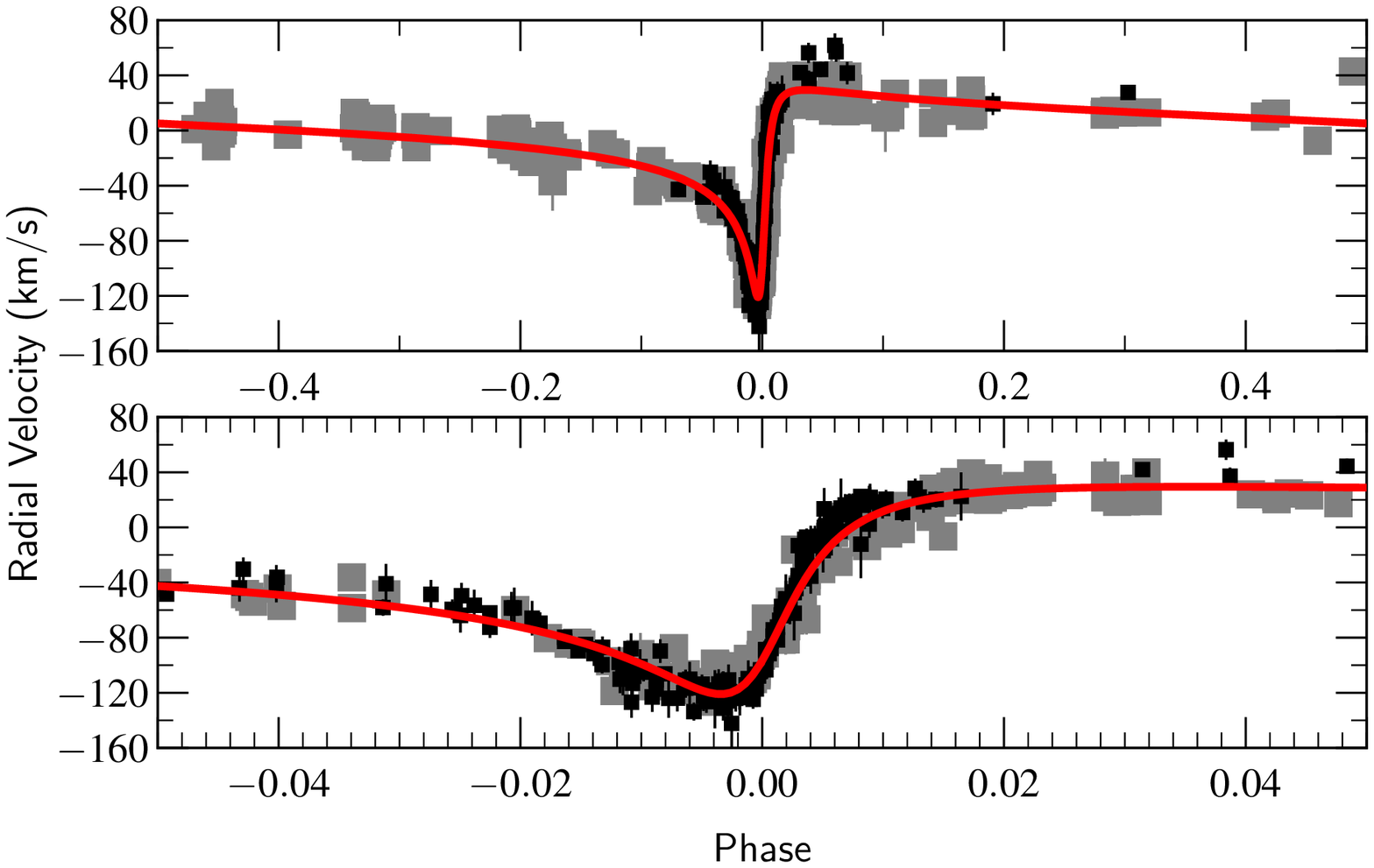}
\includegraphics[angle=0, width=8.5cm]{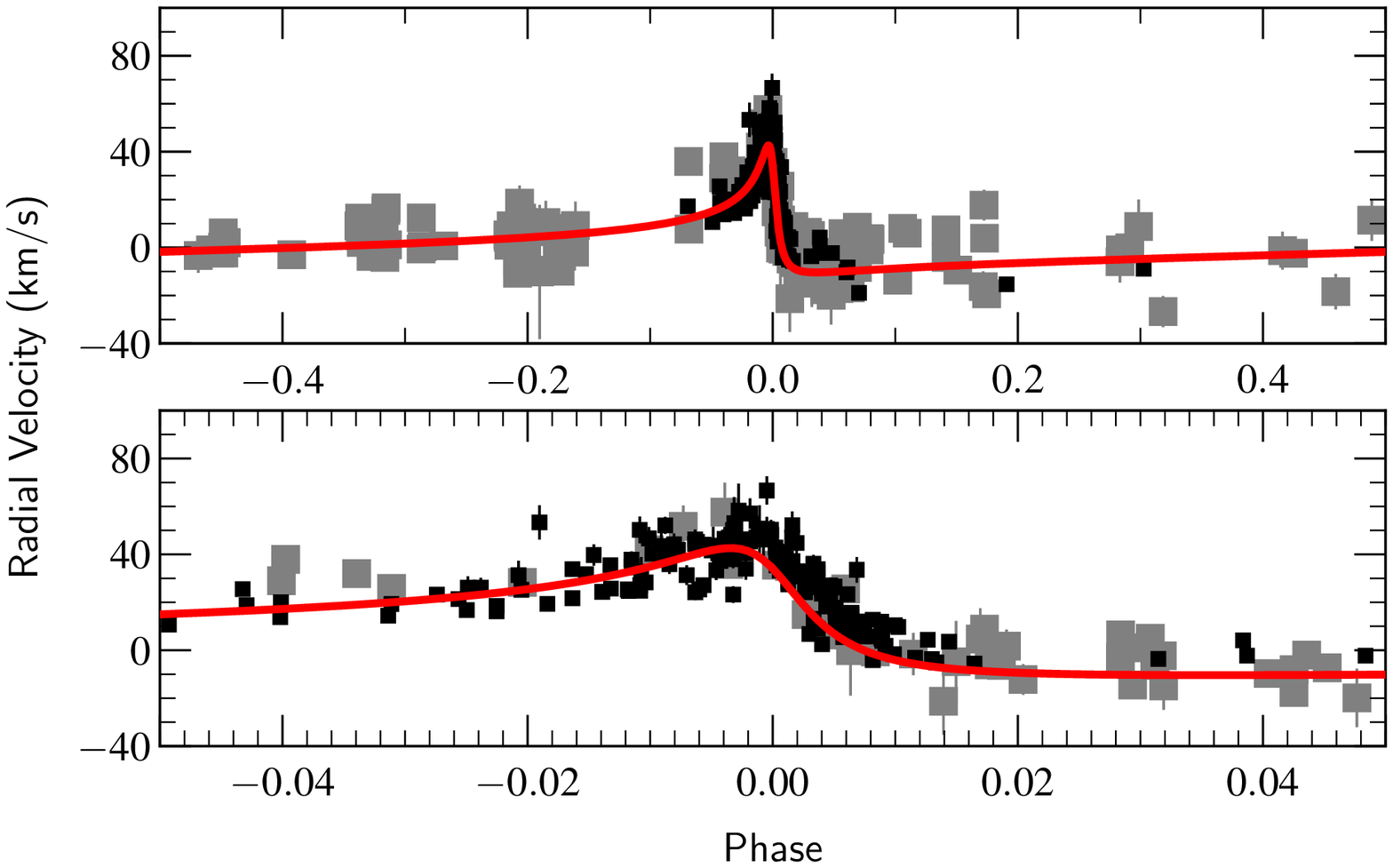}
\end{center}
\caption{All spectroscopic velocity measurements of WR\,140 with our derived
{\color{black} fit} (Table~\ref{orbitparams-table}) in red. The upper left panel shows the all the measurements for the WR component, while the upper right shows the same for the O star. The lower panels are a factor of ten magnification in the phase near periastron passage.  The plotted data include our new results (black) and historical data (grey) from \citet{2011MNRAS.418....2F} and \citet{2003ApJ...596.1295M}.}
\label{Allfit}
\end{figure*}

When we attempted to fit an orbit with measurements that had an error smaller than 5~km~s$^{-1}$, we found that the solution would have a larger $\chi^2_{\rm red}$ than our adopted orbit due to their disproportionate weighting. We then increased the error in each measurement with a small error to 5~km~s$^{-1}$ in order to fit our orbit. The visual orbit is shown in Fig.\ \ref{orbitfit} and the spectroscopic orbit with all data included is shown in the two panels of Fig.\ \ref{Allfit}.

\citet{2011ApJ...742L...1M} derived an orbital parallax for the system, which yielded a distance of 1.67$\pm0.03$~kpc. The {\it Gaia} Data Release 2 parallax (0.58$~\pm~$0.03~mas) corresponds to a distance of 1.72$\pm0.09$~kpc. However, using the work of \citet{2018AJ....156...58B}, we find that the Bayesian-inferred {\it Gaia} distance of 1.64$^{+0.08}_{-0.07}$~kpc\footnote{We also note that \citet{2020MNRAS.tmp...34R} derived a distance of 1.64$^{+0.11}_{-0.09}$ kpc using Bayesian statistics and a prior tailored for WR stars for the astrometry from {\it Gaia}.} is consistent with that of \citet{2011ApJ...742L...1M}. The Bayesian-inferred distance is preferred as it corrects for the non-linearity of the transformation and uses an expected Galactic distribution of stars, being thoroughly tested against star clusters with known distances.

{\color{black} We also note that the EDR3 data from {\it Gaia} \citep{2020arXiv201202036G} suggest a parallax of 0.5378$\pm$0.0237~mas, corresponding to a distance of 1.86$\pm$0.08 kpc, which is well outside of the allowed distances from our orbit, the {\it Gaia} DR2 distance derived by either \citet{2017yCat.2346....0B} or \citet{2020MNRAS.tmp...34R}. We speculate that this is because the EDR3 data will include data from near periastron when the photocenter seen by {\it Gaia} could shift quickly and thus interfere with excellent measurements usually given by {\it Gaia}. However, determining the actual source of the {\it Gaia} errors for WR\,140's parallax is beyond the scope of this paper. }

Our derived orbital {\color{black} parameters}, shown in lower half of Table\,~\ref{orbitparams-table}, were calculated using our derived distance in the first column.  The second column of the lower part of Table\,~\ref{orbitparams-table} shows our derived parameters calculated using the {\it Gaia} DR2 distance.  The last column of Table ~\ref{orbitparams-table} shows the results from \citet{2011ApJ...742L...1M} and \citet{2011MNRAS.418....2F} for easy reference.  We note that the distance we derive is about 2 $\sigma$ away from the accepted {\it Gaia} DR2 distance of 1.67 kpc. While this level of potential disagreement may be concerning, we also note that the recent EDR3 data for {\it Gaia} was problematic, perhaps because the measurements happened across a periastron passage. We suspect that a proper treatment of the astrometry from {\it Gaia} with the orbital motion included may solve this discrepancy, but further analysis is beyond the scope of this paper.

 We note that the masses of the O star are now lower when we allow our derived parameters to measure an orbital parallax. The mass of the WR star has a similar error as the analysis of \citet{2011ApJ...742L...1M}, but is considerably lower. In fact, we are now in a prime position to compare the system to $\gamma^2$ Velorum \citep[see the orbit presented by ][]{2017MNRAS.468.2655L}, the only other WC star with a visual orbit. $\gamma^2$ Vel's WC star has a spectral type of WC8, so is slightly cooler than the WR star in WR\,140. Its mass is $\sim 9 M_\odot$, which is only slightly less than what we infer in our orbit.

Our derived masses are lower than those derived by \citet{2011ApJ...742L...1M} with the \citet{2011MNRAS.418....2F} spectroscopic orbit when using our derived orbit without the {\it Gaia} DR2 parallax, differing by at least 3$\sigma$. However, when we take into account the {\it Gaia} DR2 parallax, the masses are within 1$\sigma$ of the values from the \citet{2011ApJ...742L...1M} analysis. The best way to solve any discrepancy in the future will be to improve the visual orbit and make use of any refinement of the {\it Gaia} parallax with future data releases.

O stars are very difficult to assign spectral types to in WR systems, due to extreme blending of the O and WR features in the optical spectrum. \citet{2011MNRAS.418....2F} found the O star to have a spectral type of O5.5fc, and the `fc' portion of the spectral type means the star should have a luminosity class of I or III \citep[e.g.,][]{2011ApJS..193...24S}. While the \citet{2011ApJ...742L...1M} solution or our solution where we adopt the {\it Gaia} distance are broadly in agreement, our derived parameters suggest that the mass is lower. If we use the O star calibrations of \citet{2005A&A...436.1049M}, then we see that the O star should have a later spectral type than given by \citet{2011MNRAS.418....2F}, although the difficulties in assigning spectral types to the companion stars in WR binaries can certainly affect this measurement.

\begin{table*}
\begin{minipage}{170mm}
\centering
\caption{ {\color{black} Orbital elements calculated using all historical data plus the new data presented in this paper are in the column ``This Work + Prior''.  In the lower half of the table we provide the derived properties of the system.  The work in this paper has two columns with values calculated from our determined distance using the visual orbit, and a second column where the parameters are calculated using the {\it Gaia} distance.} \label{orbitparams-table}}

\begin{tabular}{l c c c }
\hline \hline
Orbital Element	&	\multicolumn{2}{c}{	This Work +	}		&	Monnier 2011 + 	\\
	&	\multicolumn{2}{c}{	Prior	}		&	Fahed 2011	\\
\hline
$P$~(days)                                          	&	\multicolumn{2}{c}{	$2895.00\pm0.29$	}		&	$2896.35\pm0.20$	\\
$T_{\scalebox{.8}{$\scriptscriptstyle \rm 0$}}$~(MJD)                                      	&	\multicolumn{2}{c}{	$60636.23\pm0.53$	}		&	$46154.8\pm0.8$	\\
$e$                                             	&	\multicolumn{2}{c}{	$0.8993\pm0.0013$	}		&	   $0.8964_{\scalebox{.6}{$-0.0007$}}^{\scalebox{.6}{$+0.0004$}}$	\\
$\omega_{\scalebox{.8}{$\scriptscriptstyle \rm WR$}}~(^{\circ})$                                	&	\multicolumn{2}{c}{	$227.44\pm0.52$	}		&	$226.8\pm0.4$	\\
$K_{\scalebox{.8}{$\scriptscriptstyle \rm O$}}$~(km~s$^{-1}$)      	&	\multicolumn{2}{c}{	$26.50\pm0.48$	}		&	$30.9\pm0.6$	\\
$K_{\scalebox{.8}{$\scriptscriptstyle \rm WR$}}$~(km~s$^{-1}$)      	&	\multicolumn{2}{c}{	$-75.25\pm0.63$	}		&	$-75.5\pm0.7$	\\
$\gamma_{\scalebox{.8}{$\scriptscriptstyle \rm O$}}$~(km~s$^{-1}$)  	&	\multicolumn{2}{c}{	$3.99\pm0.37$	}		&	\ldots	\\
$\gamma_{\scalebox{.8}{$\scriptscriptstyle \rm WR$}}$~(km~s$^{-1}$) 	&	\multicolumn{2}{c}{	$0.26\pm0.32$	}		&	\ldots	\\
$i~(^{\circ})$                                 	&	\multicolumn{2}{c}{	$119.07\pm0.88$	}		&	$119.6\pm0.5$	\\
$\Omega~(^{\circ})$                               	&	\multicolumn{2}{c}{	$353.87\pm0.67$	}		&	$353.6\pm0.4$	\\
$a$ (mas)   	&	\multicolumn{2}{c}{	$8.922\pm 0.067$	}		&	$8.82\pm0.05$	\\
$\chi^{2}$                                                      	&	\multicolumn{2}{c}{	$1843.09$	}		&	\ldots	\\
$\chi^{2}_{\rm red}$  	&	\multicolumn{2}{c}{	$2.01$	}		&	\ldots	\\
\hline \hline
\multicolumn{4}{c}{Derived Properties}				\\ \hline

	&	Calculated Distance	&		{\it Gaia} Distance	&	Monnier 2011 + 	\\
	&	This Work	&		This Work	&	Fahed 2011	\\
\hline
Distance~(kpc)  	&	 $1.518\pm0.021$	&		 $1.64_{\scalebox{.6}{$-0.07$}}^{\scalebox{.6}{$+0.08$}}$	&	$1.67\pm0.03$	\\
$a$ (AU) 	&	$13.55\pm0.21$	&		$14.63\pm0.049$	&	$14.7\pm0.02$	\\
$M_{\scalebox{.8}{$\scriptscriptstyle \rm O$}}$~(M$_{\odot}$)       	&	$29.27\pm1.14$	&		$36.87\pm4.34$	&	$35.9\pm1.3$	\\
$M_{\scalebox{.8}{$\scriptscriptstyle \rm WR$}}$~(M$_{\odot}$)      	&	$10.31\pm0.45$	&		$12.99\pm1.54$	&	$14.9\pm0.5$	\\
$q = \frac{M_{\scalebox{.8}{$\scriptscriptstyle \rm WR$}}}{M_{\scalebox{.8}{$\scriptscriptstyle \rm O$}}}$                                                	&	$0.35\pm0.01$	&		$0.35\pm0.01$	&	$0.415\pm0.002$	\\
\hline	\hline

\end{tabular}
\end{minipage}
\end{table*}

\section{The Evolutionary History of WR~140}

We have attempted to understand the evolutionary history and future of WR\,140 by comparing its observational parameters to binary evolution models from the Binary Population And Spectral Synthesis (BPASS) code, v2.2.1 models, as described in detail in \citet{2017PASA...34...58E} and \citet{2018MNRAS.479...75S}. Our fitting method is based on that in \citet{2009MNRAS.400L..20E} and \citet{2011MNRAS.411..235E}. We use the $UBVJHK$ magnitudes taken from \citet{2002yCat.2237....0D}  and  \citet{2003yCat.2246....0C}. We note that the 2MASS magnitudes used here were measured in 1998, and thus were not contaminated by dust created in the 1993 IR maximum.  To estimate the extinction, we take the $V$-band magnitude from the {\color{black} BPASS} model for each time-step and compare it to the observed magnitude. If the model $V$-band flux is higher than observed, we use the difference to calculate the current value of $A_V$. If the model flux is less than observed, we assume zero extinction. We then modify the rest of the model time-step magnitudes with this derived extinction before determining how well that model fits.  Our derived value of $A_V$ is 2.4, which is in agreement with the current measurement of 2.46 \citep[e.g.,][]{1988A&A...199..217V}.
We then also require that, for an acceptable fit, the model must have a primary star that is now hydrogen free, have carbon and oxygen mass fractions that are higher than the nitrogen mass fraction and that the masses of the components and their separation match the observed values that we determine here.

The one caveat in our fitting is that the BPASS models assume circular orbits; however, as found by \citet{2002MNRAS.329..897H}, stars in orbits with the same semi-latus rectum, or same angular momentum, evolve in similar pathways independent of their eccentricity. A similar assumption was made in \citet{2009MNRAS.400L..20E}. While the orbit of WR\,140 has not circularized, we note that in cases of binary interactions within an eccentric system, the tidally-enhanced mass transfer rate near periastron can cause a perturbation in the orbit that acts to increase the eccentricity with time rather than circularize the orbit, which is a possible explanation for the current observed orbit \citep[e.g.,][]{2007ApJ...660.1624S, 2007ApJ...667.1170S, 2009ApJ...702.1387S, 2010ApJ...724..546S}.
We note that a more realistic model would require including the eccentricity. WR\,140 is clearly a system where specific modelling of the interactions may lead to interesting findings on how eccentric binaries interact.

We considered a system to be matching if the masses were $M_{WR}/M_{\odot}=10.31\pm1.99$, and $M_{O}/M_{\odot}=29.27\pm5.48$.  In selecting the period to match we use an assumption that systems with orbits that have the same semi-latus rectum are similar in their evolution. Thus taking account of the eccentricity we select models that have a separation of $log(a/R_{\odot})=2.746\pm0.1$.

Given this caveat, we find the current and initial parameters of the WR\,140 system, as presented in Table~\ref{bpass}.  The values reported in Table~\ref{bpass} are the mean values of the considered models, with error bars being the standard deviation of those models averaged.

\begin{table}
\centering
\caption{Parameters from BPASS. The primary star evolved into the current WR star.
\label{bpass}}
\begin{tabular}{l l }
\hline \hline
Initial Parameter & Value \\
\hline
$M_{\rm primary,i \rightarrow WR}$ ($M_{\odot}$)   &   $38.8\pm6.0$  \\
$M_{\rm O,i}$ ($M_{\odot}$)  &   $31.9  \pm   1.3$ \\
$\log(P_{\rm i}/d)$   &   $2.41\pm0.30$ \\
$Z$ &   $0.026\pm0.011$  \\
\hline \hline
Present Parameter & Value \\
\hline
$A(V)$  &   $2.4\pm0.2$  \\
$\log({\rm Age}/yr) $    &  $ 6.70\pm0.05 $  \\

$\log(L_{\rm primary \rightarrow WR}/L_{\odot})$  &   $5.31\pm0.06$  \\
$\log(L_{\rm O}/L_{\odot}))$  &   $5.48\pm0.04$\\

$\log(T_{\rm primary, eff \rightarrow WR}/K)$ &   $5.05\pm0.04$ \\
$\log(T_{\rm O, eff}/K)$   &   $4.43\pm0.04$ \\

\hline  \hline
\end{tabular}
\end{table}

The matching binary systems tend to interact shortly after the end of the main sequence, thus the mass transfer events occur while the primary star still has a radiative envelope. This may explain why the orbit of WR\,140 is still eccentric as deep convective envelopes are required for efficient circularization of a binary \citep{2002MNRAS.329..897H}. We also note that the mass transfer was highly non-conservative with much of the mass lost from the system. This is evident in that the orbit is significantly longer today than the initial orbit of the order of a year. The companion does accrete a few solar masses of material, so it is possible that the companion may have a significant rotational velocity.  Additionally, the companion  may be hotter than our models predict here due to the increase in stellar mass.  However, we note that the average FWHM of the He\,{\sc i} $\lambda 5876\text{\normalfont\AA}$  line {\color{black} in velocity-space} was 140~km~s$^{-1}$,  {\color{black} which if used as a proxy for the rotational velocity, $v\sin i$, } is fairly normal for young stellar clusters \citep[e.g.,][]{2006ApJ...648..580H}. If the O star rotates in the plane of the orbit, the rotational speed would be $\sim$~160~km~s$^{-1}$, slightly larger than typical O stars \citep[e.g.][]{2013A&A...560A..29R, 2015A&A...580A..92R}, but possibly less than predicted if significant accretion would have occurred \citep{2013ApJ...764..166D}.

This could also be expected if the situation is as described by \citet{2017MNRAS.464.2066S} and \citet{2018A&A...615A..65V}, where the O star's spin-up of the companion could have been braked by the brief appearance of a strong global magnetic field generated in the process \citep{2019Natur.574..211S}. Indeed, while some WR+O binaries show some degree of spin-up, that degree is observed to be much less than expected initially after accretion.

\begin{figure*}
\begin{center}
\includegraphics[angle=0,width=17cm]{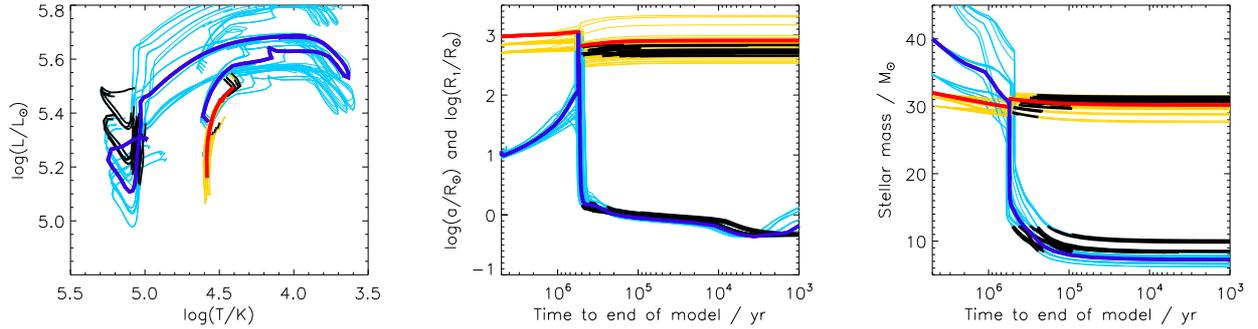}
\end{center}
\caption{{\color{black}Different aspects of evolution of the WR\,140 system are shown in these three panels. In each of the figures the blue and red bold lines represent the model with the best matching initial parameters with thinner lined models that are within the 1$\sigma$ uncertainties in initial mass, initial mass ratio, initial period and initial metallicity. Highlighted in black are the regions of the models where the mass and period of the binary match the orbital solution in this work. In the left panel we show the Hertzsprung-Russell diagram for the past and future evolution, the primary is in light/dark blue and the secondary in yellow/red. In the central panel we show the primary radius in light/dark blue and the orbital separation in yellow/red. In the right panel we show the mass of the primary in light/dark blue and the mass of the secondary in yellow/red.}}

\label{evolution}
\end{figure*}

While this discussion has used the mean values from all the BPASS models considered, we have taken the most likely fitting binary and the closest model to this and show their evolution {\color{black} as the bold curves} in Fig.~\ref{evolution}. As we describe above the interactions are modest, because the primary loses a significant amount of mass through stellar winds before mass transfer begins {\color{black} in these models}. The interaction is either a short common-envelope evolution which only shrinks the orbit slightly, or only a Roche lobe overflow with the orbit widening. In all cases the star would have become a Wolf-Rayet star without a binary interaction thus making the interactions modest since most mass loss was done via stellar winds.

The most confusing thing about WR\,140 is the significantly low estimated age of only 5.0~Myrs ($\log({\rm Age}/yr) = 6.70 $). There are relatively few other stars in the volume of space near WR\,140 that would be members of a young cluster. It is therefore a good example of how sometimes clusters may form one very massive star rather than a number of lower-mass stars. The location of the stellar wh$\rm \bar{a}$nau\footnote{The M$\rm\bar{a}$ori word for extended family.} is an open question in its history. It is difficult to make this system older, even if we assume that the Wolf-Rayet star could have been the result of evolution in a triple system and the result of a binary merger. Indeed, such a scenario would not explain how such a massive O star like the companion star could exist. Its presence sets a hard upper limit on the age of the system of approximately 5.0~Myrs.

\section{Conclusions}

We have presented an updated set of orbital elements for WR\,140, using newly acquired spectroscopic and interferometric data combined with previously published measurements.
{\color{black}We simultaneously fit all data to produce our orbital solution, and derived masses from our orbit of $M_{\rm WR} = 10.31\pm0.45 M_\odot$ and $M_{\rm O} = 29.27\pm1.14 M_{\odot}$.  We noted in our discussion that the O star mass seems a bit low given an earlier spectral classification, but that classification of O stars in WR systems is challenging.}
Future measurements of more WR binaries will be crucial to test stellar models.
{\color{black} For WR\,140, a detailed spectral model of the binary, as done for other WR binaries resolved with interferometry \citep[e.g.,][]{2016MNRAS.461.4115R} would allow for the derived parameters of the system to be used to constrain the models of WR stars and their winds.
}

We also discussed the possible evolutionary history of the system in comparison to the BPASS models. The results show that the majority of the envelope is lost by stellar winds with binary interactions only removing a modest amount of material. The measurements presented here should allow for more precise comparisons with the stellar evolutionary and wind models for massive (binary) stars in the future. Furthermore, these results will be used as a foundation for interpretation of multiple data sets that have been collected, including the X-ray variability (Corcoran et al., in prep) and wind collisions \citep{2021MNRAS.503..643W}. While these orbital elements are well defined, future interferometric observations with MIRCX will allow for exquisite precision in new measurements, along with additional spectroscopic observational campaigns during periastron passages. MIRCX imaging at the times closest to periastron could pinpoint the location of the dust formation in the system, which could be observable in November 2024.

\section*{Acknowledgements}

This work is based in part upon observations obtained with the Georgia State University Center for High Angular Resolution Astronomy Array at Mount Wilson Observatory.  The CHARA Array is supported by the National Science Foundation under Grant No. AST-1636624 and AST-1715788.  Institutional support has been provided from the GSU College of Arts and Sciences and the GSU Office of the Vice President for Research and Economic Development. Some of the time at the CHARA Array was granted through the NOAO community access program (NOAO PropID: 17A-0132 and 17B-0088; PI: Richardson). MIRC-X received funding from the European Research Council (ERC) under the European Union's Horizon 2020 research and innovation programme (grant No.\ 639889).  This work has made use of data from the European Space Agency (ESA) mission {\it Gaia} (\url{https://www.cosmos.esa.int/gaia}), processed by the {\it Gaia}
Data Processing and Analysis Consortium (DPAC, \url{https://www.cosmos.esa.int/web/gaia/dpac/consortium}). Funding for the DPAC has been provided by national institutions, in particular the institutions participating in the {\it Gaia} Multilateral Agreement.

NDR acknowledges previous postdoctoral support by the University of Toledo and by the Helen Luedtke Brooks Endowed Professorship, along with NASA grant \#78249.  HS acknowledges support from the European Research Council (ERC) under the European Union's DLV-772225-MULTIPLES Horizon 2020 and from the FWO-Odysseus program under project G0F8H6. JDM acknowledges support from AST-1210972, NSF 1506540 and NASANNX16AD43G.
AFJM is grateful to NSERC (Canada) for financial aid. PMW is grateful to the Institute for Astronomy for continued hospitality and access to the facilities of the Royal Observatory. {\color{black}This research made use of Astropy,\footnote{http://www.astropy.org} a community-developed core Python package for Astronomy \citep{astropy:2013, astropy:2018}.}

\section*{Data Availability}

All measurements used in this analysis are tabulated either in this paper or included in cited references.




\bibliographystyle{mnras}
\bibliography{WR140spec}






\appendix

\section{Radial Velocity Measurements}

\begin{table}
\caption{Measured radial velocities	for	the	new	spectra	presented in this paper.
\label{RadialVelocities}}
\begin{tabular}{c	c	c	l	}
\hline	\hline
HJD$-2450000.5$ &	WR Velocity &	O Velocity &	Source	\\
	&(km/s)	&(km/s)&	\\
\hline
4645.59101	&	$	-42.8	$		$\pm$	5.5	&	$	17.2	$	$\pm$	1.7	&	ESPaDOnS	\\
4703.46011	&	$	-48.3	$		$\pm$	3.2	&	$	10.7	$	$\pm$	1.1	&	ESPaDOnS	\\
4755.28504	&	$	-58.2	$		$\pm$	5.9	&	$	14.4	$	$\pm$	1.3	&	ESPaDOnS	\\
5722.38773	&	$	27.6	$		$\pm$	2.9	&	$	-8.9	$	$\pm$	1.0	&	ESPaDOnS	\\
7615.9085	&	$	-44.0	$		$\pm$	9.8	&	$	25.5	$	$\pm$	2.0	&	Leadbeater	\\
7616.82776	&	$	-30.3	$		$\pm$	8.5	&	$	18.8	$	$\pm$	1.2	&	Stober	\\
7624.76623	&	$	-40.8	$		$\pm$	13.5	&	$	13.8	$	$\pm$	1.8	&	Ozuyar	\\
7624.91809	&	$	-36.2	$		$\pm$	5.1	&	$	20	$	$\pm$	2.4	&	Garde	\\
7651.01573	&	$	-40.9	$		$\pm$	14.4	&	$	19.3	$	$\pm$	2.7	&	Thomas	\\
7661.76968	&	$	-48.6	$		$\pm$	10.4	&	$	23.2	$	$\pm$	2.5	&	Ozuyar	\\
7666.89338	&	$	-59.5	$		$\pm$	7.3	&	$	21.4	$	$\pm$	2.7	&	Guarro	\\
7668.83826	&	$	-63.8	$		$\pm$	12.4	&	$	16.8	$	$\pm$	2.2	&	Guarro	\\
7669.09369	&	$	-49.6	$		$\pm$	9.4	&	$	26.1	$	$\pm$	4.7	&	Thomas	\\
7672.14171	&	$	-56.6	$		$\pm$	11.4	&	$	26.1	$	$\pm$	4.1	&	Thomas	\\
7675.86058	&	$	-62.0	$		$\pm$	2.4	&	$	16.2	$	$\pm$	2.0	&	Campos	\\
7675.89578	&	$	-72.4	$		$\pm$	7.8	&	$	18.5	$	$\pm$	2.2	&	Guarro	\\
7681.06131	&	$	-58.5	$		$\pm$	11.6	&	$	31.4	$	$\pm$	6.0	&	Thomas	\\
7681.7328	&	$	-58.4	$		$\pm$	14.6	&	$	25.2	$	$\pm$	3.1	&	Ozuyar	\\
7685.99396	&	$	-66.0	$		$\pm$	12.5	&	$	53.3	$	$\pm$	7.1	&	Thomas	\\
7687.88062	&	$	-69.5	$		$\pm$	7.8	&	$	19.4	$	$\pm$	2.4	&	Guarro	\\
7693.78032	&	$	-82.7	$		$\pm$	4.9	&	$	33.9	$	$\pm$	2.6	&	Leadbeater	\\
7693.78366	&	$	-79.7	$		$\pm$	5.2	&	$	21.7	$	$\pm$	2.5	&	Guarro	\\
7697.01575	&	$	-89.4	$		$\pm$	4.1	&	$	31.7	$	$\pm$	3.1	&	Lester	\\
7698.83037	&	$	-85.0	$		$\pm$	4.7	&	$	39.7	$	$\pm$	4.5	&	Guarro	\\
7700.83225	&	$	-91.5	$		$\pm$	9.0	&	$	24.4	$	$\pm$	2.7	&	Guarro	\\
7702.75581	&	$	-99.5	$		$\pm$	6.4	&	$	35.5	$	$\pm$	2.5	&	Leadbeater	\\
7702.87022	&	$	-87.1	$		$\pm$	7.8	&	$	25.7	$	$\pm$	2.9	&	Guarro	\\
7706.85286	&	$	-97.9	$		$\pm$	11.3	&	$	25.4	$	$\pm$	2.6	&	Guarro	\\
7707.0665	&	$	-110.1	$		$\pm$	10.1	&	$	24.7	$	$\pm$	3.3	&	Thomas	\\
7707.74176	&	$	-109.3	$		$\pm$	12.6	&	$	37.8	$	$\pm$	3.4	&	Leadbeater	\\
7707.77422	&	$	-98.4	$		$\pm$	5.4	&	$	34.8	$	$\pm$	3.6	&	Guarro	\\
7709.69196	&	$	-88.0	$		$\pm$	11	&	$	50.2	$	$\pm$	5.6	&	Ozuyar	\\
7709.81017	&	$	-126.8	$		$\pm$	11.3	&	$	27.2	$	$\pm$	4.7	&	Ribeiro	\\
7709.81296	&	$	-101.8	$		$\pm$	8.2	&	$	24.8	$	$\pm$	2.6	&	Guarro	\\
7710.04092	&	$	-113.1	$		$\pm$	9.5	&	$	32.7	$	$\pm$	6.1	&	Thomas	\\
7711.07536	&	$	-104.4	$		$\pm$	5.0	&	$	28.3	$	$\pm$	5.0	&	Thomas	\\
7711.84949	&	$	-101.0	$		$\pm$	12.6	&	$	46.7	$	$\pm$	4.7	&	Leadbeater	\\
7712.70276	&	$	-105.0	$		$\pm$	6.7	&	$	40.2	$	$\pm$	3.1	&	Leadbeater	\\
7714.75764	&	$	-123.0	$		$\pm$	10.8	&	$	43.5	$	$\pm$	6.2	&	Ribeiro	\\
7715.71599	&	$	-117.4	$		$\pm$	5.8	&	$	52.1	$	$\pm$	3.5	&	Leadbeater	\\
7715.73729	&	$	-110.7	$		$\pm$	5.4	&	$	39.2	$	$\pm$	3.0	&	Beradi	\\
7716.69472	&	$	-89.7	$		$\pm$	8.6	&	$	35.9	$	$\pm$	4.0	&	Ozuyar	\\
7717.79801	&	$	-106.2	$		$\pm$	5.3	&	$	44.2	$	$\pm$	4.4	&	Guarro	\\
7718.71246	&	$	-124.1	$		$\pm$	14.1	&	$	41.8	$	$\pm$	4.9	&	Garde	\\
7720.77745	&	$	-123.9	$		$\pm$	9.6	&	$	31.4	$	$\pm$	4.1	&	Ribeiro	\\
7722.74233	&	$	-110.8	$		$\pm$	7.5	&	$	44.4	$	$\pm$	4.5	&	Guarro	\\
7722.77181	&	$	-116.7	$		$\pm$	6.8	&	$	46.2	$	$\pm$	4.1	&	Beradi	\\
7722.80655	&	$	-111.7	$		$\pm$	6.7	&	$	24.2	$	$\pm$	3.3	&	Campos	\\
7723.74669	&	$	-113.0	$		$\pm$	15.3	&	$	25.4	$	$\pm$	2.9	&	Campos	\\
7723.75284	&	$	-110.2	$		$\pm$	5.6	&	$	43.1	$	$\pm$	4.3	&	Guarro	\\
7724.74938	&	$	-116.3	$		$\pm$	3.8	&	$	44	$	$\pm$	4.4	&	Guarro	\\
7724.75337	&	$	-133.7	$		$\pm$	6.4	&	$	27.1	$	$\pm$	3.1	&	Campos	\\
7726.68947	&	$	-113.7	$		$\pm$	6.8	&	$	41.6	$	$\pm$	4.4	&	Garde	\\
7727.06099	&	$	-122.6	$		$\pm$	10.6	&	$	43.6	$	$\pm$	7.9	&	Thomas	\\
7727.76572	&	$	-126.4	$		$\pm$	8.2	&	$	33.4	$	$\pm$	4.1	&	Ribeiro	\\
7728.75588	&	$	-121.2	$		$\pm$	7.1	&	$	37.2	$	$\pm$	4.3	&	Ribeiro	\\
7729.72349	&	$	-118.0	$		$\pm$	3.2	&	$	46.4	$	$\pm$	4.9	&	Guarro	\\
7729.79283	&	$	-124.7	$		$\pm$	21.1	&	$	44.8	$	$\pm$	6.3	&	Campos	\\
7730.68289	&	$	-113.7	$		$\pm$	10.7	&	$	43.6	$	$\pm$	5.3	&	Ozuyar	\\

\hline  \hline
\end{tabular}
\end{table}

\begin{table}
\contcaption{Measured radial velocities	for	the	new	spectra	presented in this paper.}
\begin{tabular}{c	c	c	l	}
\hline	\hline
HJD$-2450000.5$ &	WR	Velocity &	O	Velocity &	Source	\\
	&(km/s)	&(km/s)&	\\
\hline
7730.73046	&	$	-123.2	$		$\pm$	8.2	&	$	47	$	$\pm$	5	&	Guarro	\\
7731.69913	&	$	-119.2	$		$\pm$	6.4	&	$	50.2	$	$\pm$	3.8	&	Beradi	\\
7731.74248	&	$	-119.5	$		$\pm$	7.1	&	$	38	$	$\pm$	3.8	&	Guarro	\\
7731.7559	&	$	-131.2	$		$\pm$	9.1	&	$	41.2	$	$\pm$	5	&	Ribeiro	\\
7731.76974	&	$	-111.5	$		$\pm$	7.1	&	$	23.2	$	$\pm$	3.4	&	Campos	\\
7732.00337	&	$	-122.5	$		$\pm$	9.7	&	$	53	$	$\pm$	10.9	&	Thomas	\\
7732.6973	&	$	-123.5	$		$\pm$	9.8	&	$	44.8	$	$\pm$	5.8	&	Garde	\\
7732.89655	&	$	-125.9	$		$\pm$	14.5	&	$	46.1	$	$\pm$	4.1	&	Leadbeater	\\
7733.04298	&	$	-110.7	$		$\pm$	11.2	&	$	58.2	$	$\pm$	11.3	&	Thomas	\\
7733.78171	&	$	-142.2	$		$\pm$	29.9	&	$	46.3	$	$\pm$	8.8	&	Campos	\\
7734.74934	&	$	-119.8	$		$\pm$	6.2	&	$	40.6	$	$\pm$	3.9	&	Guarro	\\
7734.75611	&	$	-124.0	$		$\pm$	9.5	&	$	34	$	$\pm$	4.6	&	Ribeiro	\\
7735.69391	&	$	-123.5	$		$\pm$	8.8	&	$	45.4	$	$\pm$	4.8	&	Garde	\\
7735.73915	&	$	-122.6	$		$\pm$	13.8	&	$	57	$	$\pm$	6.3	&	Guarro	\\
7737.75114	&	$	-109.7	$		$\pm$	13.2	&	$	50.7	$	$\pm$	5.4	&	Guarro	\\
7737.99405	&	$	-112.2	$		$\pm$	6.5	&	$	50.4	$	$\pm$	10.1	&	Thomas	\\
7738.92489	&	$	-124.7	$		$\pm$	6.9	&	$	46.2	$	$\pm$	7.7	&	Thomas	\\
7739.69769	&	$	-117.2	$		$\pm$	13	&	$	66.6	$	$\pm$	5.9	&	Leadbeater	\\
7739.72805	&	$	-106.5	$		$\pm$	7.9	&	$	50.6	$	$\pm$	5.1	&	Guarro	\\
7740.70156	&	$	-104.4	$		$\pm$	7.9	&	$	50.3	$	$\pm$	4.1	&	Beradi	\\
7740.72612	&	$	-103.3	$		$\pm$	4.6	&	$	48.2	$	$\pm$	4.8	&	Guarro	\\
7741.75243	&	$	-102.9	$		$\pm$	10.5	&	$	42.9	$	$\pm$	6.5	&	Ribeiro	\\
7741.93606	&	$	-88.6	$		$\pm$	4.7	&	$	39	$	$\pm$	4	&	Lester	\\
7741.95381	&	$	-103.3	$		$\pm$	6.9	&	$	41.2	$	$\pm$	8	&	Thomas	\\
7741.96782	&	$	-89.3	$		$\pm$	8.3	&	$	35.4	$	$\pm$	3.6	&	Lester	\\
7741.99769	&	$	-92.7	$		$\pm$	9.4	&	$	40.3	$	$\pm$	4.1	&	Lester	\\
7742.75752	&	$	-91.5	$		$\pm$	7.2	&	$	33.3	$	$\pm$	4.1	&	Ribeiro	\\
7743.22505	&	$	-87.8	$		$\pm$	7.8	&	$	40.8	$	$\pm$	2.8	&	ESPaDOnS	\\
7743.70079	&	$	-78.9	$		$\pm$	9.4	&	$	39.5	$	$\pm$	3	&	Beradi	\\
7743.75725	&	$	-74.3	$		$\pm$	12.5	&	$	35.3	$	$\pm$	3.9	&	Guarro	\\
7744.75494	&	$	-81.6	$		$\pm$	6.6	&	$	28.7	$	$\pm$	3.5	&	Ribeiro	\\
7744.76281	&	$	-72.4	$		$\pm$	6.6	&	$	27.4	$	$\pm$	3	&	Guarro	\\
7745.70176	&	$	-63.4	$		$\pm$	8.4	&	$	47.2	$	$\pm$	3.9	&	Beradi	\\
7745.72762	&	$	-57.4	$		$\pm$	10.9	&	$	52.1	$	$\pm$	5.8	&	Guarro	\\
7745.74729	&	$	-64.0	$		$\pm$	8.7	&	$	37.2	$	$\pm$	4.3	&	Campos	\\
7746.74793	&	$	-55.1	$		$\pm$	5.5	&	$	44.7	$	$\pm$	4.9	&	Guarro	\\
7746.76355	&	$	-51.8	$		$\pm$	17.7	&	$	33	$	$\pm$	4.9	&	Campos	\\
7747.75535	&	$	-53.4	$		$\pm$	16.9	&	$	30.9	$	$\pm$	4.4	&	Garde	\\
7748.70943	&	$	-46.8	$		$\pm$	11.4	&	$	26.2	$	$\pm$	2.4	&	Beradi	\\
7748.72376	&	$	-47.3	$		$\pm$	13.8	&	$	24.2	$	$\pm$	2.6	&	Guarro	\\
7748.75753	&	$	-62.3	$		$\pm$	19.7	&	$	31.6	$	$\pm$	4.6	&	Ribeiro	\\
7749.70118	&	$	-39.0	$		$\pm$	13.9	&	$	27.5	$	$\pm$	2.3	&	Beradi	\\
7749.72075	&	$	-34.9	$		$\pm$	12.7	&	$	28.3	$	$\pm$	2.9	&	Guarro	\\
7749.75724	&	$	-13.2	$		$\pm$	10.1	&	$	6.8	$	$\pm$	1	&	Ribeiro	\\
7750.69524	&	$	-37.3	$		$\pm$	12.6	&	$	35.8	$	$\pm$	2.8	&	Leadbeater	\\
7750.71981	&	$	-22.7	$		$\pm$	11.4	&	$	36.1	$	$\pm$	3.7	&	Guarro	\\
7750.75891	&	$	-25.3	$		$\pm$	6.1	&	$	13.1	$	$\pm$	1.5	&	Ribeiro	\\
7751.70041	&	$	-24.2	$		$\pm$	7.8	&	$	32.1	$	$\pm$	3.9	&	Garde	\\
7751.72145	&	$	-19.2	$		$\pm$	11.5	&	$	33.8	$	$\pm$	3.4	&	Guarro	\\
7751.75795	&	$	-7.1	$		$\pm$	6.2	&	$	8.9	$	$\pm$	1.2	&	Ribeiro	\\
7751.75888	&	$	-28.9	$		$\pm$	12.3	&	$	26.3	$	$\pm$	3.6	&	Campos	\\
7752.69799	&	$	-13.2	$		$\pm$	7.7	&	$	23.6	$	$\pm$	2.9	&	Garde	\\
7752.7211	&	$	-7.7	$		$\pm$	8.9	&	$	25.3	$	$\pm$	2.6	&	Guarro	\\
7752.73721	&	$	-35.4	$		$\pm$	12.7	&	$	2.5	$	$\pm$	0.8	&	Campos	\\
7753.72269	&	$	-8.1	$		$\pm$	9.2	&	$	18.2	$	$\pm$	1.8	&	Guarro	\\
7754.69082	&	$	-6.4	$		$\pm$	13	&	$	27.1	$	$\pm$	3.2	&	Garde	\\
7754.7034	&	$	-9.8	$		$\pm$	10.2	&	$	22.5	$	$\pm$	1.6	&	Beradi	\\
7754.94876	&	$	-7.3	$		$\pm$	15.5	&	$	26.1	$	$\pm$	3.6	&	Thomas	\\
7755.70308	&	$	-16.8	$		$\pm$	15.9	&	$	26.6	$	$\pm$	2.3	&	Leadbeater	\\

\hline \hline
\end{tabular}
\end{table}

\begin{table}
\contcaption{Measured radial velocities	for	the	new	spectra	presented in this paper.}
\begin{tabular}{c	c	c	l	}
\hline	\hline
HJD$-2450000.5$ &	WR	Velocity &	O	Velocity &	Source	\\
	&(km/s)	&(km/s)&	\\
\hline
7755.94031	&	$	-2.2	$		$\pm$	11.7	&	$	15.8	$	$\pm$	1.6	&	Lester	\\
7755.9461	&	$	13.4	$		$\pm$	15.2	&	$	17.7	$	$\pm$	2.3	&	Thomas	\\
7755.963	&	$	0	$		$\pm$	10.2	&	$	10.5	$	$\pm$	1.2	&	Lester	\\
7756.74935	&	$	1.3	$		$\pm$	9.9	&	$	14.8	$	$\pm$	1.8	&	Guarro	\\
7757.70237	&	$	3.8	$		$\pm$	11.2	&	$	8.3	$	$\pm$	1	&	Beradi	\\
7757.70469	&	$	-8.3	$		$\pm$	12.2	&	$	13.3	$	$\pm$	1.2	&	Leadbeater	\\
7757.72899	&	$	8.5	$		$\pm$	5.1	&	$	5.5	$	$\pm$	0.9	&	Guarro	\\
7758.72732	&	$	5.1	$		$\pm$	10.1	&	$	23.3	$	$\pm$	2.5	&	Guarro	\\
7759.69877	&	$	10.1	$		$\pm$	11.2	&	$	15.7	$	$\pm$	1.8	&	Garde	\\
7759.72275	&	$	9.2	$		$\pm$	8.3	&	$	12.3	$	$\pm$	1.4	&	Guarro	\\
7759.76482	&	$	-2.7	$		$\pm$	2.9	&	$	15	$	$\pm$	2	&	Ribeiro	\\
7759.97778	&	$	14	$		$\pm$	21.4	&	$	6.3	$	$\pm$	1.4	&	Thomas	\\
7760.73924	&	$	11.9	$		$\pm$	3.4	&	$	7.5	$	$\pm$	1	&	Guarro	\\
7760.95617	&	$	16.3	$		$\pm$	7.7	&	$	33.6	$	$\pm$	5.3	&	Thomas	\\
7761.95875	&	$	15.5	$		$\pm$	9.1	&	$	12	$	$\pm$	1.3	&	Lester	\\
7762.74853	&	$	18.3	$		$\pm$	6.3	&	$	11.8	$	$\pm$	1.4	&	Guarro	\\
7762.76955	&	$	6.3	$		$\pm$	5.1	&	$	5.8	$	$\pm$	1	&	Ribeiro	\\
7764.7039	&	$	16.1	$		$\pm$	11.1	&	$	5.8	$	$\pm$	0.8	&	Beradi	\\
7764.72541	&	$	22.5	$		$\pm$	4.9	&	$	12.7	$	$\pm$	1.5	&	Guarro	\\
7764.73166	&	$	-12.1	$		$\pm$	24.7	&	$	-4.1	$	$\pm$	1	&	Campos	\\
7766.72684	&	$	20.7	$		$\pm$	9.4	&	$	5	$	$\pm$	0.8	&	Guarro	\\
7766.74306	&	$	2.5	$		$\pm$	10	&	$	11.9	$	$\pm$	1.4	&	Leadbeater	\\
7766.94453	&	$	19.7	$		$\pm$	11.9	&	$	3.9	$	$\pm$	0.7	&	Lester	\\
7766.96052	&	$	21.8	$		$\pm$	7.3	&	$	5.5	$	$\pm$	1	&	Thomas	\\
7767.73057	&	$	18.5	$		$\pm$	8.5	&	$	1.9	$	$\pm$	0.7	&	Guarro	\\
7769.73312	&	$	15.2	$		$\pm$	7.3	&	$	-1.6	$	$\pm$	0.7	&	Guarro	\\
7769.94296	&	$	13.8	$		$\pm$	7.3	&	$	10.4	$	$\pm$	1.2	&	Lester	\\
7770.75387	&	$	20.4	$		$\pm$	6.8	&	$	9.7	$	$\pm$	1.3	&	Guarro	\\
7774.71494	&	$	10.8	$		$\pm$	6.4	&	$	-3.2	$	$\pm$	0.8	&	Leadbeater	\\
7777.73525	&	$	28.3	$		$\pm$	7.1	&	$	4.4	$	$\pm$	0.8	&	Guarro	\\
7778.71117	&	$	22.1	$		$\pm$	6.1	&	$	-3.7	$	$\pm$	0.7	&	Beradi	\\
7779.72978	&	$	18.5	$		$\pm$	8	&	$	-5.0	$	$\pm$	0.8	&	Leadbeater	\\
7782.7386	&	$	20.3	$		$\pm$	5.9	&	$	3.6	$	$\pm$	0.8	&	Leadbeater	\\
7788.73838	&	$	22.4	$		$\pm$	17.4	&	$	-5.6	$	$\pm$	0.9	&	Leadbeater	\\
7832.19508	&	$	42	$		$\pm$	4.9	&	$	-3.6	$	$\pm$	0.8	&	Guarro	\\
7852.19978	&	$	56.3	$		$\pm$	7.3	&	$	4.1	$	$\pm$	0.9	&	Thomas	\\
7853.13919	&	$	37.1	$		$\pm$	6.3	&	$	-2.3	$	$\pm$	0.7	&	Guarro	\\
7881.13262	&	$	44.4	$		$\pm$	5.4	&	$	-2.3	$	$\pm$	0.7	&	Guarro	\\
7915.14428	&	$	61.8	$		$\pm$	8.7	&	$	-10.3	$	$\pm$	1.4	&	Thomas	\\
7918.07741	&	$	57.2	$		$\pm$	7.4	&	$	-8.3	$	$\pm$	1.2	&	Thomas	\\
7944.85441	&	$	41.7	$		$\pm$	8.1	&	$	-18.9	$	$\pm$	2.5	&	Guarro	\\
8293.62071	&	$	19.3	$		$\pm$	8.1	&	$	-15.3	$	$\pm$	1.2	&	ESPaDOnS	\\
\hline \hline
\end{tabular}
\end{table}



\bsp	
\label{lastpage}
\end{document}